\documentclass[final,1p,times,twocolumn]{elsarticle}
\usepackage{times,amsmath,amssymb,amsthm}
\usepackage{bm}
\usepackage{amsfonts} 
\usepackage{subfigure}
\usepackage{mathtools}
\usepackage{overcite}
\usepackage{footnpag}
\usepackage{cleveref}
\makeatletter
\def\@author#1{\g@addto@macro\elsauthors{\normalsize%
    \def\baselinestretch{1}%
    \upshape\authorsep#1\unskip\textsuperscript{%
      \ifx\@fnmark\@empty\else\unskip\sep\@fnmark\let\sep=,\fi
      \ifx\@corref\@empty\else\unskip\sep\@corref\let\sep=,\fi
      }%
    \def\authorsep{\unskip,\space}%
    \global\let\@fnmark\@empty
    \global\let\@corref\@empty  %% Added
    \global\let\sep\@empty}%
    \@eadauthor={#1}
}
\usepackage{soul}

\begin{document}
\begin{frontmatter}
\title{Optimal Trajectories for Propellant-Free Rendezvous Missions}
\author[Toyota]{Mohamed Shouman \corref{cor1}}
\ead{Mohamed.Shouman@Toyota-ti.ac.jp}
\author[UCSD,SDSU1]{Ahmed Atallah}
\ead{aatallah@eng.ucsd.edu}
\author[UCSD,SDSU2]{Mohammad S. Ramadan}
\ead{msramada@eng.ucsd.edu}

\cortext[cor1]{Corresponding author}
\address[Toyota]{Research center of the smart vehicle, Toyota Technological Institute (TTI), Nagoya, Aichi, Japan},

\address[UCSD]{Department of Mechanical \&\ Aerospace Engineering, University of California, San Diego, La Jolla CA 92093-0411, USA.} 

\address[SDSU1] {Department of Aerospace Engineering, San Diego State University, San Diego, La Jolla CA, USA.} 

\address[SDSU2] {Electrical and Computer Engineering Department, San Diego State University, San Diego, La Jolla CA, USA.} 

\begin{abstract}
The paper provides a new approach to utilizing space environmental forces in time- and energy-optimal, propellant-less spacecraft rendezvous missions. Considering the nonlinear form of the relative dynamic equations, rendezvous missions are posed as optimal control problems subject to input saturation. We conduct a direct optimal control approach to obtain optimal trajectories and control inputs. 
Initially, we consider the differential drag only and conduct a comprehensive analysis of the effect of altitude on the required control input and achieved cost function. Lorentz forces are then utilized with the differential drag, reducing the time required for time-optimal missions.  
For energy-optimal missions with combined differential drag and Lorentz forces, a weighting matrix in the cost function is introduced to adjust the relative contributions of these forces.
\end{abstract}
\begin{keyword}
Rendezvous Mission, Differential Atmospheric Drag, Lorentz Forces, Trajectory Optimization.
\end{keyword}
\end{frontmatter}
\section{Introduction}

The new era of space debris removal and formation flying missions require innovative ideas for trajectory tracking problems.
The surge of interest in decreasing fuel consumption in Earth orbiter spacecraft missions creates the need to exploit space environmental forces, such as atmospheric drag, Lorentz forces, and solar radiation pressure, in various orbit control applications. 

A pioneering work on incorporating the differential atmospheric drag to orbital control was investigated by Leonard et al. \cite{Leonard1989} in 1989. In their work, a linear quadratic regulator (LQR) is implemented to control a linearized form of the relative dynamics for rendezvous missions. Subsequently, various control algorithms have been developed to use differential atmospheric drag in different orbit control applications, especially in satellite formation flying (SFF) missions \cite{Cho2016, Mazal2016,  Shouman2016}. 

In SFF missions, the coupling between the radial and the tangential directions in the dynamic models facilitates the control of both in-plane components. However, the normal to the orbit plane movement is deemed uncontrollable, as previously studied by several authors 
\cite{Leonard1989, Bevilacqua2010, Reid2011, Kumar2007IEEE, Varma2012, Shouman2019d, Shouman2021}. Despite the lack of controllability, Ivanov et al. \cite{IVANOV2018} proposed a simple LQR to control the relative dynamic model, using only the atmospheric forces, including the lift component. Therefore, the atmospheric drag as an exogenous disturbance, originally, is now part of the construction of the control problem and hence, a decision variable in the optimal control problem. However, the physical realization of this control law imposes geometrical and design considerations, which are impractical for most satellite designs.

Towards more exploits of the space environment, the utilization of the Lorentz forces for orbital control is continuously being investigated. In 2005, Peck et al. \cite{Peck2005} proposed a simple linear time-invariant (LTI) model for orbit control missions of Earth escape and drag compensation.
Later, the viability of using Lorentz forces for formation flying and rendezvous missions has been extensively studied \cite{Sobiesiak2015, Huang2015, Vatankhanghadim2017}. 
Pollock et al. \cite{Pollock2011} provided an approximate analytical solution to the relative motion equations, in which the Lorentz force accounts for a significant component of spacecraft propulsion.

The viability of Lorentz forces for SFF missions has been confirmed by Tsujii et al. \cite{Tsujii2013}, using relative dynamics states for circular and elliptic orbit configurations. 
Sobiesiak and Damaren analyzed the controllability of a formation flying mission using a perturbed relative dynamics model based on the Brouwer formulation \cite{Brouwer1959} of the mean orbital element secular drift rates. 
Sobiesiak et al. \cite{Sobiesiak2015} proved that stabilizing the uncontrolled states can be substantiated in the Lyapunov sense for specific orbit configurations.

As an advanced step, Vatankhanghadim and Damaren used a hybrid LQR scheme based on the relative states \cite{Sobiesiak2015_2} for SFF missions in elliptical orbits using the combined control actions of Lorentz forces and impulsive thrusters \cite{Vatankhanghadim2017}. 
Subsequently, Shouman et al. \cite{Shouman2021} analyzed the controllability of linearized systems with several combinations of space environmental forces, including atmospheric drag and Lorentz forces, for orbit control applications.
In his following paper, Shouman \cite{Shouman2020IAC} has proposed an extensive analysis of the boundaries for the finite-time reachability closures of the relative dynamics models using only bounded differential atmospheric force by using two approaches. 
The first estimates the optimum solutions of Hamilton-Jacobi (HJ) for the nonlinear models by maximizing the general performance index of the optimization problem. The second approach exploits the eigenstructure of the linearized Hills-Clohessy-Wiltshire (HCW) model for circular orbits \cite{Hills1878, Clohessy1960}.

This paper focuses on solving time- and energy-optimal control problems for rendezvous missions involving nonlinear relative dynamics.
The control input is restricted to using the differential atmospheric drag with or without minor support by the Lorentz forces.
In considerable measure, optimal feedback control laws are desirable due to their stability and feasibility guarantees. 
However, acquiring such control laws requires solving a Hamilton-Jacobi-Bellman (HJB) equation which is intractable due to the so-called curse of dimensionality. 
Instead, we appeal to Pontryagin's maximum principle, which drops the requirement of a closed-loop solution, and therefore avoids the curse of dimensionality. 
Although optimal trajectories are solved in an open-loop sense, feedback can be established through a receding horizon. 
This technique is shown to yield asymptotic stability and recursive feasibility, under specific terminal conditions, over the infinite horizon. This approach presents the foundational core result of Model Predictive Control (MPC) due to Keerthi and Gilbert \cite{keerthi1988optimal}; properties of the open-loop finite-horizon can yield features of the infinite horizon optimal feedback control. 
The contribution of the paper three-fold, which are: 

\begin{itemize}
    \item In contrast to many references \cite{Mazal2016, Shouman2016, Bevilacqua2010, Reid2011, Shouman2021}, in which they implement differential drag for only the in-plane motion or combine drag with thrusts to achieve full controllability for the whole relative dynamics, our paper investigates energy- and time-optimal trajectories for the entire dynamic models using the differential atmospheric drag only.
    The full controllability of the dynamic using the atmospheric drag follows from the representation of the follower satellite's relative velocity vector in the leader satellite's coordinate system.  
    \item It comprehensively analyzes the altitude effects on mean control actions and the cost function for the minimum-energy approach. 
    \item It derives a practical solution for handling differential drag drawbacks with minor support from Lorentz forces for both minimum-energy and minimum-time approaches.  
    
\end{itemize}

The rest of the paper is organized as follows. \Cref{Sec:2} presents the nonlinear relative dynamics equations and the differential atmospheric drag and Lorentz forces in the radial--tangential--normal (RTN) coordinate system. In \cref{Sec:4}, the formulations of the optimal control problems for the propellant-free spaces rendezvous missions are introduced. The numerical results for the control with the atmospheric drag only and a combination of the atmospheric drag and Lorentz forces are divided into two separate subsections in \cref{sec:5}. \Cref{sec:6} states the conclusions and recommendations to enhance the performance of the proposed methods.

\section{Dynamics Models}\label{Sec:2}
In an inverse gravity field, the equations of the motion for the leader satellite are given by 
\begin{equation} \label{eq:3}
\begin{split}
\ddot{r}_c=r_c \dot{\theta}^2-\frac{\mu}{r_c^2},\quad
\ddot{\theta}=- 2 \frac{\dot{r}_c \dot{\theta}}{r_c^2}
\end{split}
\end{equation}
where $r_c$ is the radial distance between the leader satellite and the center of the Earth, and $\theta$ is its argument of latitude.  
The relative motion between the leader and follower satellites is described in the radial--tangential--normal (RTN) coordinate system as follows:
\cite{Vallado2013}
\begin{equation}\label{eq:2}
\begin{split}
\ddot{x}&=x \dot{\theta}^2+ y \ddot{\theta} + 2\dot{y} \dot{\theta} -\frac{\mu(r_c+x)}{r^3}+\frac{\mu}{r_c^2}+ a_x \\ 
\ddot{y}&=y \dot{\theta}^2- x \ddot{\theta} - 2\dot{x} \dot{\theta} -\frac{\mu y}{r^3}+ a_y \\ 
\ddot{z}&=-\frac{\mu z}{r^3}+ a_z
\end{split}
\end{equation} 
where $x, y,$ and $z$ are the relative positions in the radial, tangential, and normal directions, respectively; $r$ is the radial distance of the follower satellite from the center of the Earth, i.e., $r =\sqrt{(r_c+x)^2+y^2+z^2}$, and $\mathbf{a}  \triangleq \begin{bmatrix}a_{x}& a_{y}&a_{z}\end{bmatrix}^T$ represents the difference in space environmental forces between the leader and follower satellites.

%\section{Space Environmental Forces} \label{sec:3}

We regard the differential atmospheric drag and Lorentz forces as the control inputs, hence, the acceleration is
\begin{equation} 
\label{e:envirom_forces}
\mathbf{a}=\mathbf{a}_d+\mathbf{a}_l
\end{equation}
where $\mathbf{a}_d$ and $\mathbf{a}_l$ are their corresponding accelerations, respectively. Later in the numerical results section, the differential drag component is solely considered at first, then with the inclusion of that of Lorentz forces.

\subsection{Differential Atmospheric Drag} \label{sec:3a}
The atmospheric drag represents the most significant non-gravitational force acting on LEO satellites. 
The magnitude of acceleration required to counteract aerodynamic drag decreases dramatically with altitude.
Therefore, using aerodynamic drag to control formation flying is viable only for LEO satellites. 
The control action of the atmospheric drag is generated from the difference between the leader and follower satellites' atmospheric drag perturbations. 
This atmospheric drag difference is represented by
\begin{equation}\label{eq:drag}
\begin{split}
\bm{a}_d&=\bm{a}_{d_f}-\bm{a}_{d_l}\\
& = -\frac{1}{2} \rho C_d \frac{A_f}{m_f} \bm{v}_{rel_{f}} ||\bm{v}_{rel_{f}}|| +\frac{1}{2} \rho C_d \frac{A_l}{m_l} \bm{v}_{rel_{l}} ||\bm{v}_{rel_{l}}||\\
\end{split} 
\end{equation}
where $C_d$ is the drag coefficient of both satellites, $\rho$ is the aerodynamic density, and $\bm{a}_{d_l}$ and $\bm{a}_{d_f}$ are the aerodynamic drag force vectors for the leader and follower satellites, respectively. 
The relative velocity of the leader satellite in the RTN coordinate system  $\bm{{v}}_{rel_l}$ is given by

\begin{equation}\label{eq:121}
\bm{v}_{rel_l} = \left[ \begin{array}{c} 
\dot{r}_c
\\ r_c (\dot{\theta} -\omega_e \cos i) \\  r_c \omega_e \sin i \cos \theta \end{array} \right]
\end{equation}
%Our study estimates a solution to the control action for the generic formation flying problem.
and the relative velocity of the follower satellite in the leader coordinate system \cite{huang2015pseudospectral} 
\begin{equation}\label{eq:113}
\bm{v}_{rel_f} = \left[ \begin{array}{c} 
\dot{r}_c+\dot{x}-y(\dot{\theta}-\omega_e \cos{i})-z\omega_e \sin{i} \cos{\theta}
\\ \dot{y}+(r_c+x)(\dot{\theta}-\omega_e \cos{i})+z\omega_e \sin{i} \sin{\theta} \\ \dot{z}+(r_c+x)\omega_e \sin{i}\cos{\theta}-y\omega_e \sin{i} \sin{\theta} \end{array} \right]
\end{equation}
where $i$ is the inclination of the leader satellite, $\omega_e= 7.292 \times 10^{-5}$ sec$^{-1}$ is the angular velocity of Earth's rotation around its axis.

 In \cref{eq:drag}, $\frac{A_l}{m}$ and $\frac{A_f}{m}$ denote the areas over mass ratio for the leader and follower satellites, respectively. The difference between these values generates the differential drag force. 
 We assume using rotating drag plates for both satellites:
 \begin{equation}\label{eq:2s}
\begin{split}
\frac{A_l}{m}=\frac{A}{m} \sin{(\alpha +\delta \alpha)}
,\quad
\frac{A_f}{m}=\frac{A}{m} \sin{(\alpha -\delta \alpha)}
\end{split}
\end{equation}
where $\frac{A}{m}$ and $\alpha$ are the nominal drag area over mass and drag plate angle, respectively, and $\delta \alpha$ is the difference in the drag plate angle. 
  
\subsection{Lorentz Forces} \label{Sec:3b}
The Lorentz force is generated by a charged satellite moving through a geomagnetic field, whereas the satellite is assumed as a charged point of mass rotating in the Terrestrial magnetic field \cite{Tsujii2013}. 
The Lorentz force acting on a satellite is given by
\begin{equation}\label{eq:112}
\bm{a}_{l} =\frac{q}{m} \bm{v}_{rel} \times \bm{B}
\end{equation}
Here, the fraction $\frac{q}{m}$ is the charge-to-mass ratio of the satellite, and $\bm{v}_{rel}$ is the relative velocity for the follower satellite given by \cref{eq:113}. 
The geomagnetic field of Earth $\bm{B}$ is given for a perfect tilted dipole moment as \cite{Shouman2021}
\begin{equation}\label{eq:lorentz}
\bm{B} = \frac{B_0}{r_c^3} \left[ 3( \hat{\bm{n}} \cdot \bm{\hat{r}_c}) \bm{\hat{r}_c} - \hat{\bm{n}} \right]
\end{equation}
where $B_0$ is the strength of the field in weber kilometer $ (B_0= 8\times 10^{15})$, $\bm{\hat{n}}$ is the magnetic dipole unit vector, and $\bm{\hat{r}_c}$ is the unit vector of the inertial position vector of the leader satellite. This formula shall be transformed from the inertial coordinate system to the RTN coordinate system. 
Pollock et al. \cite{Pollock2010} derived a simple representation for the geomagnetic field vector in the RTN coordinate system:
\begin{equation}\label{eq:120}
\bm{B} = \frac{B_0}{r^3} \left[ \begin{array}{c}  -2 \sin \lambda_n (\cos \beta \cos \theta + \sin \beta \cos i \sin \theta) - 2 \cos \lambda_n \sin i \sin \theta \\
\sin \lambda_n (\sin \beta \cos i \cos \theta - \cos \beta \sin \theta ) + \cos \lambda_n \sin i \cos \theta \\
\cos \lambda_n \cos i - \sin \lambda_n \sin \beta \sin i \end{array} \right]
\end{equation}
where $\beta$ is defined by $\beta =\omega_e t- \Omega$ and $\lambda_n$ is the co-latitude angle between the geographic north pole and geomagnetic pole, ($\lambda_n=11.5$ deg). 

\section{Problem Formulation} \label{Sec:4} 

The system in \cref{eq:2} defines a state \begin{align*}
\mathbf{x}(t)\triangleq\begin{bmatrix}x(t)& y(t)&z(t)&\dot{x}(t)&\dot{y}(t)& \dot{z}(t)\end{bmatrix}^T
\end{align*}  and
a control input $\mathbf{u}\triangleq \begin{bmatrix}\delta \alpha \end{bmatrix}$. 
Rewrite \eqref{eq:3} and \eqref{eq:2} as
\begin{align}
\label{eq:state}
\dot {\mathbf{x}}(t)=\mathbf{f}(\mathbf{x}(t),\mathbf{x}_l(t),\mathbf{u}(t))
\end{align}
where $\mathbf{x}_l(t)$ is the vector of the states of the leader satellite, i.e., % $\mathbf{x}_l(t)$ is given by
\begin{equation}
    \mathbf{x}_l(t)= \begin{bmatrix} r_c(t)&\theta(t)&\dot{r}_c(t)&\dot{\theta}(t)\end{bmatrix}^T
\end{equation}

Our objective is to compute energy- and time-optimal trajectories for rendezvous missions. 
These optimal control problems can be posed as follows. 
Determine the states, $\mathbf{x}(t)$, and the control inputs, $\mathbf{u}(t)$, that minimize the performance index
\begin{equation} \label{eq:energy}
\begin{split}
J= \int_{t_0}^{t_f}\mathcal{L}(t,\mathbf{x}(t),\mathbf{u}(t)) \text{d} t
\end{split}
\end{equation}
 subject to the dynamics constraints in 
 \cref{eq:state} and \cref{eq:2}, %,eq:3,e:envirom_forces},
 and constraints on the control inputs:
 \begin{equation}
\begin{split}
  |\mathbf{u}(t)|\leq \mathbf{u}_{\text{max}}
 \end{split}
 \end{equation}
 and the boundary conditions  
 \begin{equation}
\begin{split}
     \mathbf{x}(t_0) &= \mathbf{x}_0\\
     \mathbf{x}_l(t_0) &= \mathbf{x}_{l_0}\\
     \mathbf{x}(t_f) &= \mathbf{0}     
        \end{split}
 \end{equation}
In the minimum-energy problem, 
\begin{equation} \label{eq:energy2}
\mathcal{L} (t,\mathbf{x}(t),\mathbf{u}(t))= \mathbf{u}(t)^TR\mathbf{u}(t)
\end{equation}
for some $R>0$, while in the minimum-time problem
\begin{equation} \label{eq:time}
\begin{split}
\mathcal{L} (t,\mathbf{x}(t),\mathbf{u}(t))=  {1 }
\end{split}
\end{equation}
Using the CasADi toolbox \cite{andersson2019casadi}, we employed the multiple shooting method to transcribe the above optimal control problems into finite-dimensional approximations, resulting in nonlinear programming problems.  
The IPOPT \cite{wachter2006implementation} are then implemented to solve the nonlinear programming problems. 

\section{Numerical Results} \label{sec:5} 
Previous works \cite{Leonard1989, Varma2012, Shouman2021} show that the system in \cref{eq:2} with the differential atmospheric drag is controllable for the in-plane motion only. In the first subsection, we analyze the ability to control the in-plane and out-of-plane components of the relative dynamic equations with the differential atmospheric drag alone.
Then, the following subsection additionally integrates the Lorentz forces with differential atmospheric drag. Each subsection presents the results for energy-optimal and time-optimal cases. 
Further analyzes are conducted to study the impact of the altitude.

For the leader and follower satellites, the nominal drag area over mass $A/m = 0.1 \ $ m$^2$/kg is assumed with a nominal drag plate angle of $\alpha =\pi/4$, respectively. 
The orbit parameters of the leader satellite are given in Table \ref{tab:1}. 
The equivalent altitude above the Earth's surface is $h_{ellp}= 400$ km.
It is difficult to determine an exact value of the density in the upper aerodynamic layers, and many international standards attempt to promote one density model over another by specifying numerous parameters to select the best model for a particular mission \cite{Vallado2014, Perez2016}. Among these models, we estimate the density based on the simple exponential density model (CIRA 72) \cite{Vallado2013}.
\begin{equation}\label{eq:density}
\rho =\rho_0 e^{\big(-\frac{h_{ellp}-h_0}{H}\big)} 
\end{equation}
\noindent
where $h_0$,  $\rho_0$, and $H$ are the actual altitude, base attitude, nominal density at the base attitude, and scale height, respectively \cite{Vallado2013}. The density at this altitude equals $ 3.7250 \times 10^{-12}$ according to \cref{eq:density}. 
The initial position and velocity difference for all cases are stated as follows:
 \begin{equation}
     \mathbf{x}(t_0) = \begin{bmatrix} 0&1000 &100 &0 & 0 & 0\end{bmatrix}^T
 \end{equation}
For the differential atmospheric drag alone,
\begin{equation}
    \mathbf{u}_\text{max} = \begin{bmatrix} 
\pi/4 &0.0
    \end{bmatrix}^T
\end{equation}
and when considering atmospheric drag and Lorentz forces
\begin{equation}
    \mathbf{u}_\text{max} = \begin{bmatrix} 
\pi/4 &0.01
    \end{bmatrix}^T
\end{equation}

\begin{table} [tb]
\caption{Orbit Parameters of the Leader Satellite} \label{tab:1}
\center{
\begin{tabular}{p{4cm}p{2cm}p{2cm}p{2cm}}
\hline\hline
Parameter &Symbol & Unit & Value \\\hline
Semi-Major axis & $a$ & km & $6778$ \\
Inclination & $i$ & deg & $30$ \\
Eccentricity & $e$ & $--$& $0.001$ \\
RAAN &$\Omega$ &  deg  & $10$\\
Argument of Perigee &$\omega$ &  deg  & $10$\\True Anomaly &$\nu$ &  deg  & $10$\\\hline\hline
\end{tabular}
}
\end{table} 

\subsection{Control with Atmospheric Drag}\label{sec:5.1}

Designing optimal trajectories for the whole relative dynamical model with differential drag alone was not presented to the best of the author's knowledge. In this subsection, we offer the simulation results for only the atmospheric differential drag for the minimum energy and optimal time control problems.  

\begin{figure}[tb]
\centering
\subfigure[Position]
{\label{fig:1a}
\includegraphics[clip, trim=0cm 0cm 1cm 1cm, width=.48\textwidth]{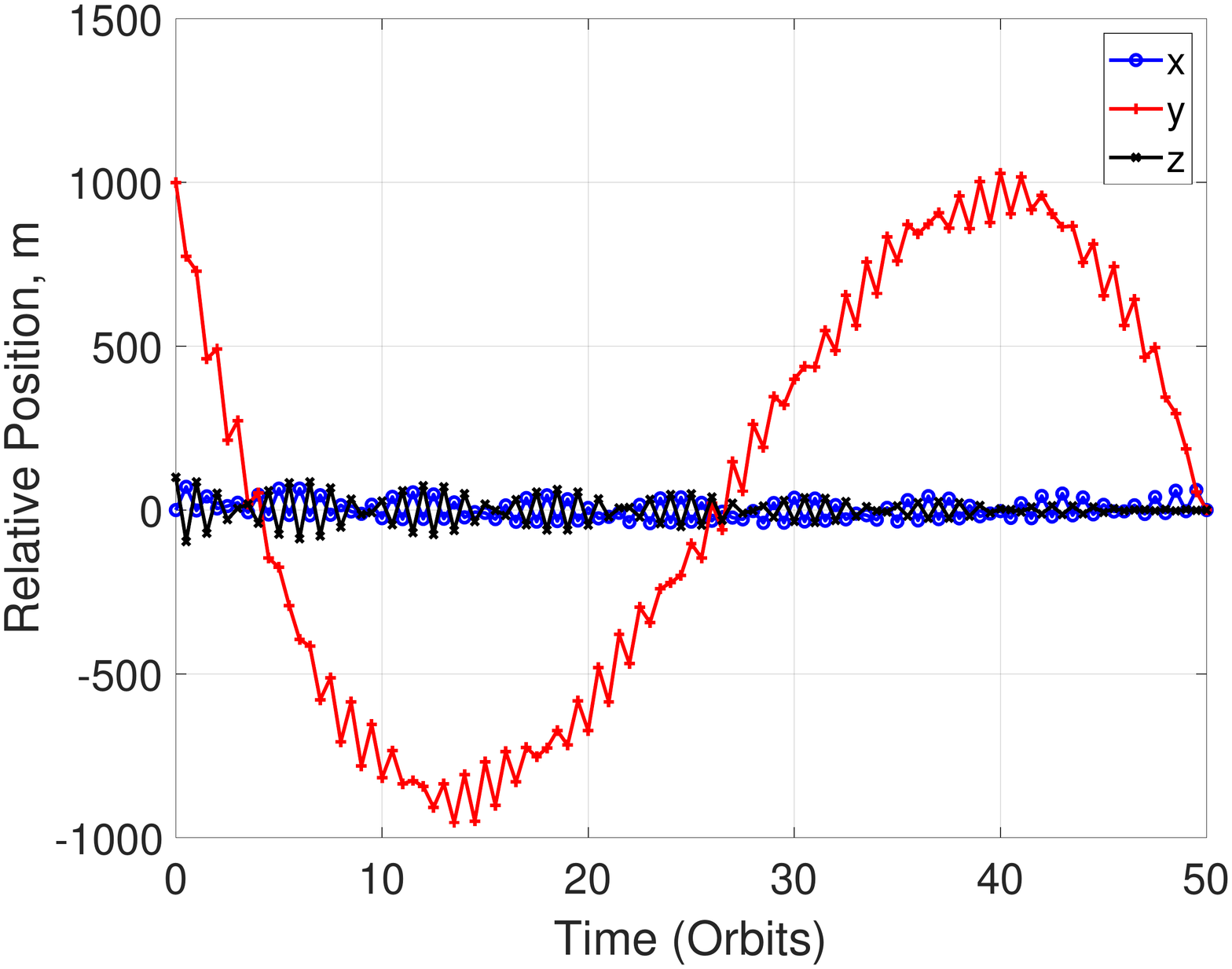}}
\subfigure[Velocity]
{\label{fig:1b}
\includegraphics[clip, trim=0cm 0cm 1cm 1cm, width=.48\textwidth]{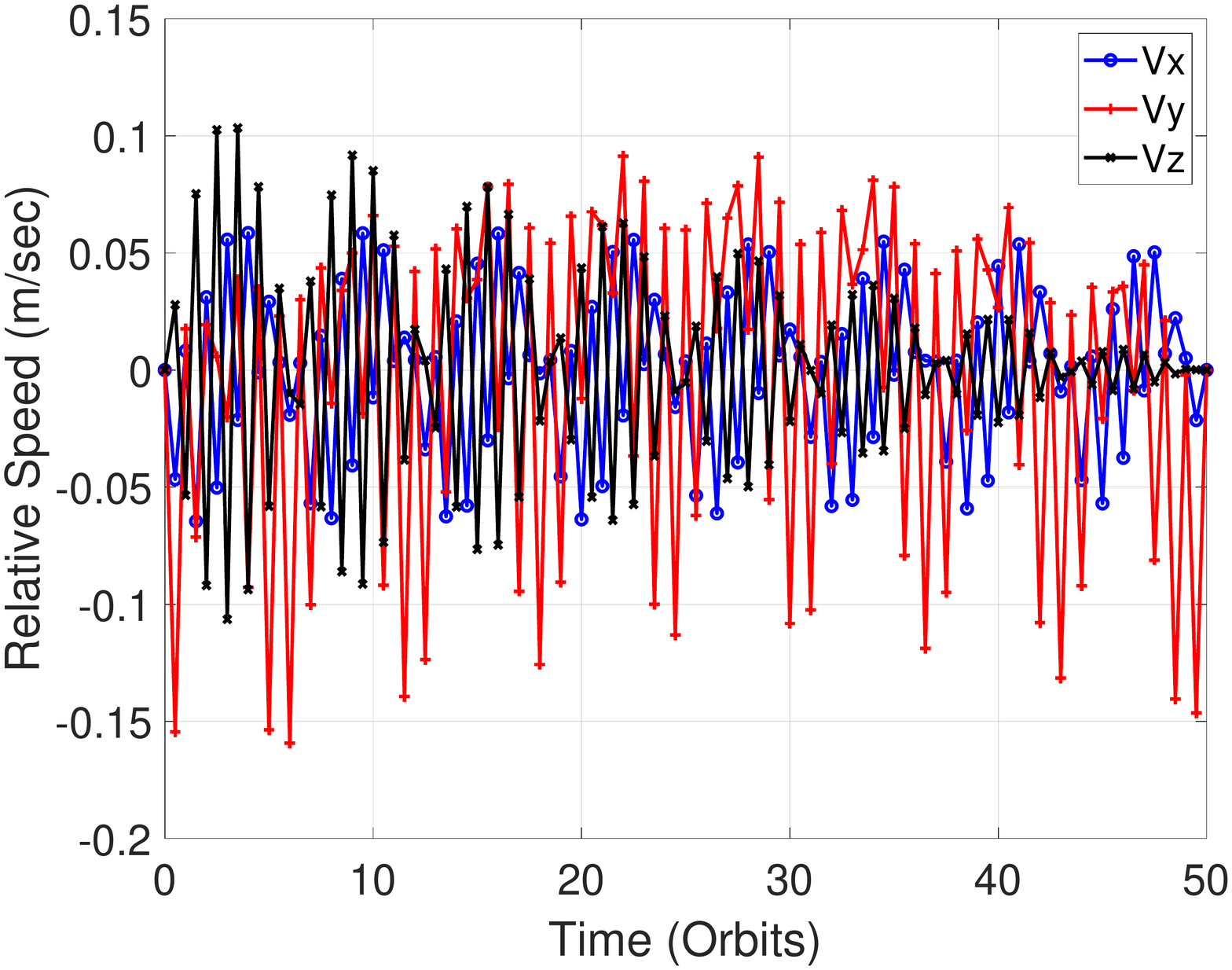}}
\caption{Error in Position and Velocity For Minimum energy Approach (Drag).}
\label{fig:1}
\end{figure}
\begin{figure}[tb]
\centering
{\includegraphics[clip, trim=0cm 0cm 0cm 0cm, width=.50 \textwidth]{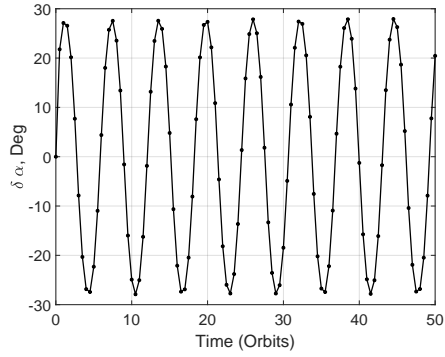}}
\caption{Control action for Minimum Energy Approach (Drag).}
\label{fig:2}
\end{figure}
\noindent
\begin{enumerate} 
\item Minimum Energy.

This part presents the minimum-energy optimal control problem results only when considering the differential atmospheric drag. \Cref{fig:1a} shows the time history for the radial and tangential direction, while \Cref{fig:1b} depicts the time history for the velocity components.
In \cref{fig:2}, the trajectory of the control input $(\delta \alpha)$ is illustrated with final time $t_f= 50$ orbital periods, approximately equal to $3.5$ days. The control action of the atmospheric drag plate angle $\delta \alpha$ changes to higher than $25$ deg. The total cost function for this scenario $J = 1.88 \times 10^{4}$ with weighting matrix $R=I$ in \cref{eq:energy2}. 
\begin{figure}[tb]
\centering
\subfigure[Mean Control Actions]
{\label{fig:3na}
\includegraphics[clip, trim=0cm 0cm 1cm 1cm, width=.48\textwidth]{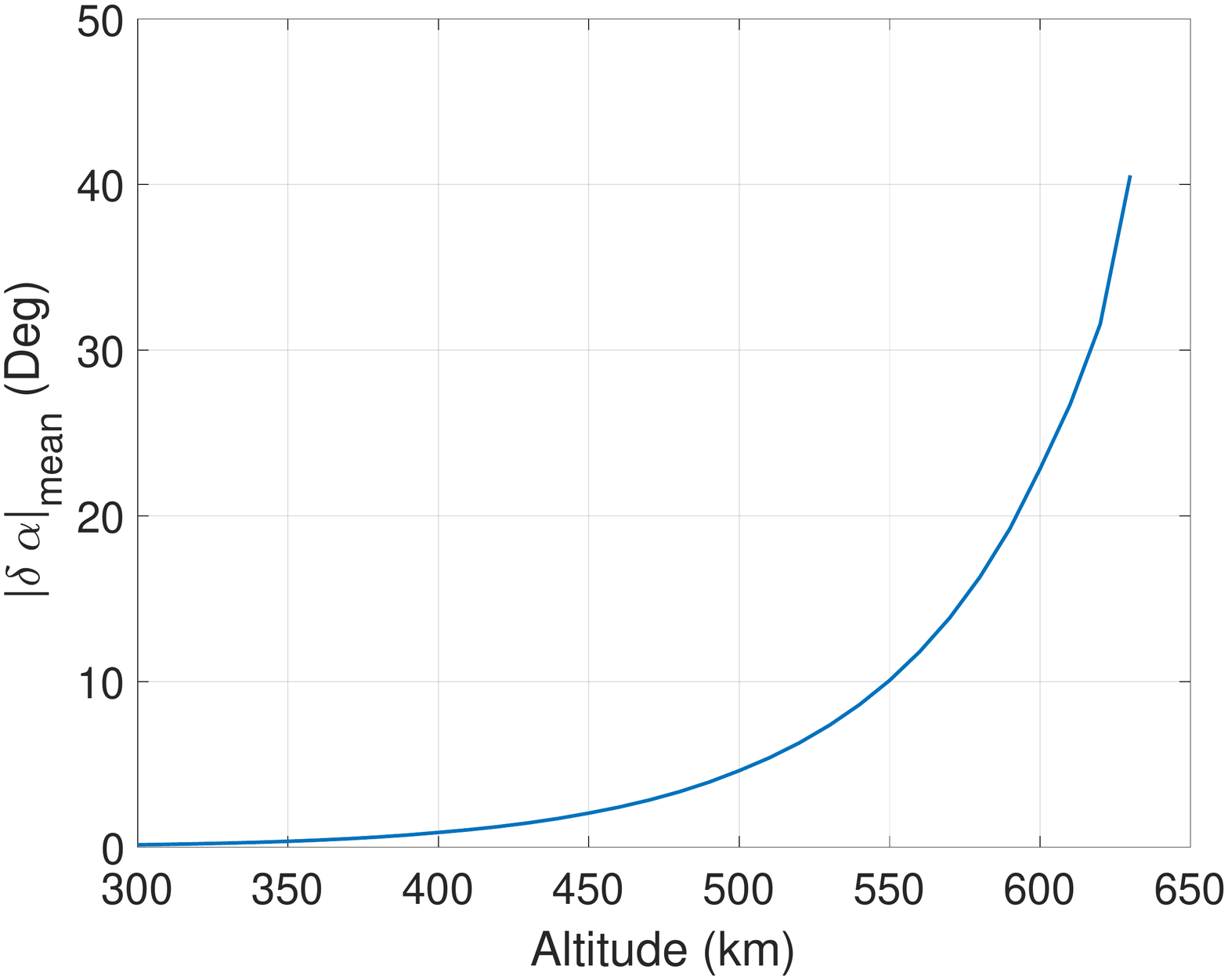}}
\subfigure[Total Cost Function]
{\label{fig:3nb}
\includegraphics[clip, trim=0cm 0cm 1cm 1cm, width=.48\textwidth]{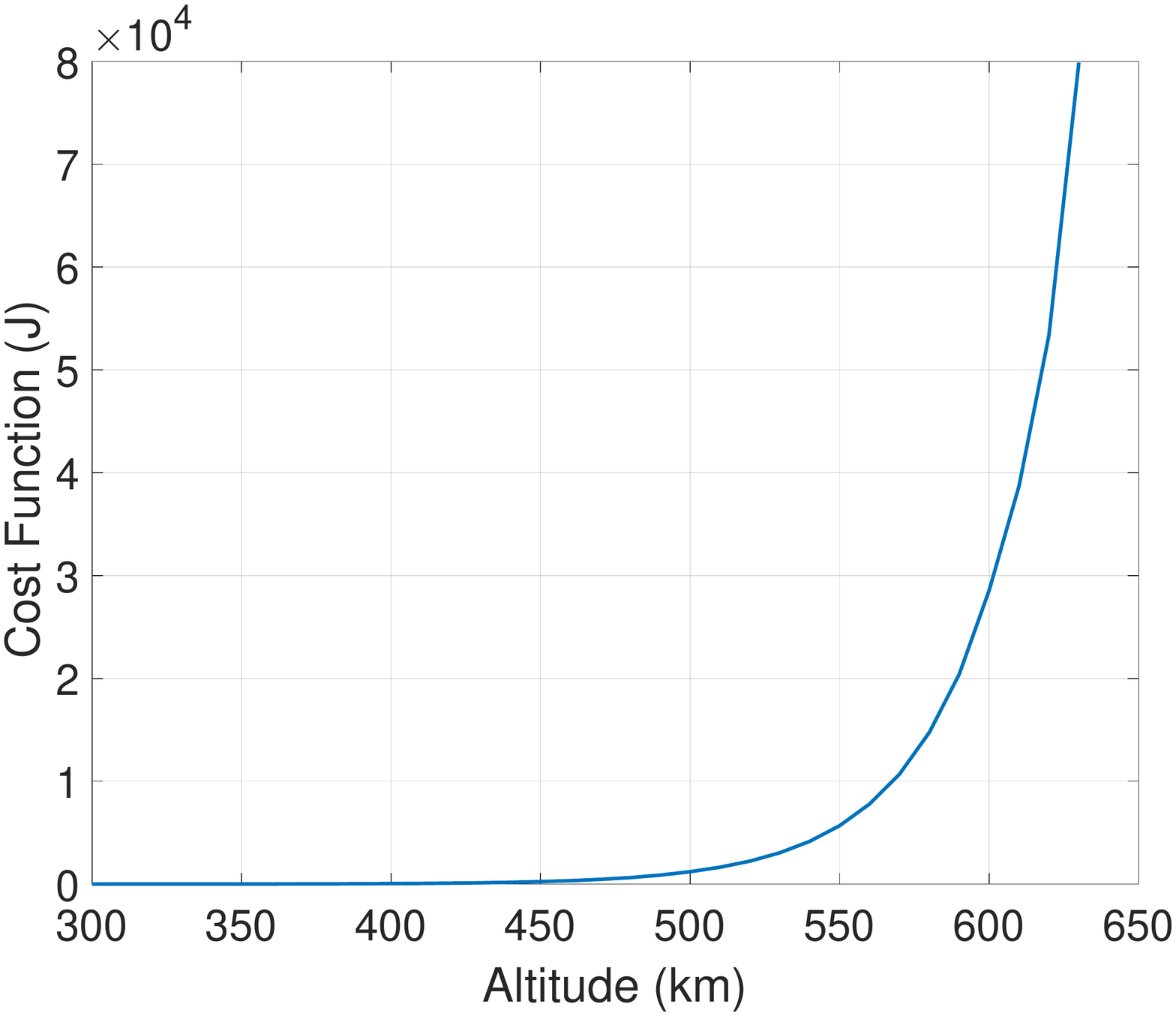}}
\caption{Mean Control Input and Total Cost Function with Various Altitudes $h_{ellp}$ (Drag).}
\label{fig:3n}
\end{figure}
\noindent

The atmospheric drag depends on density value and, consequently, on the satellite altitude. Therefore, we need to analyze the effect of altitude on the cost function for the minimum-energy problem. \Cref{fig:3n} presents the change of the mean control action $|\delta \alpha|_{mean}$ and the cost function $J$ of differential atmospheric drag with $t_f=1000$ orbital periods, which is approximately equal to 70 days. The mission duration is selected to achieve the minimum requirements for most cases of various altitudes from $h_{ellp}=300$ km to $h_{ellp}= 650$ km with atmospheric density ranging from $\rho= 2.418 \times 10^{-11}$ kg/m $^3$ to $\rho= 7.249 \times 10^{-14}$ kg/m $^3$. It is shown in \cref{fig:3na} that the mean absolute value of drag plate angle difference over the whole duration increases dramatically from  $0.14 $ deg at $h_{ellp} =300$ km to $41 $ deg at $h_{ellp} =630$ km. At the same time, atmospheric drag alone can't achieve this mission for higher altitudes within the given duration and satellite parameters. 
\Cref{fig:3nb} illustrates that the
cost function increases dramatically with a small change in altitude, ranging from $J \approx 1.11 $ at $h_{ellp} =300$ km to $J \approx 8 \times 10^4 $ at $h_{ellp} =630$ km.    

\item Minimum Time

\Cref{fig:33} shows the time histories of the relative positions and velocity components for the minimum-time approach, respectively. It is illustrated that the mission requires approximately $177000$ sec or $29.5$ orbits with realistic values for drag plate areas over mass ratios to be implemented with differential drag alone. 
The minimum-time trajectory optimization yields control trajectories that exhibit approximately the bang-bang property, as shown in \cref{fig:44}.

\begin{figure}[tb]
\centering
\subfigure[Position]
{\label{fig:33a}
\includegraphics[clip, trim=0cm 0cm 1cm 1cm, width=.48\textwidth]{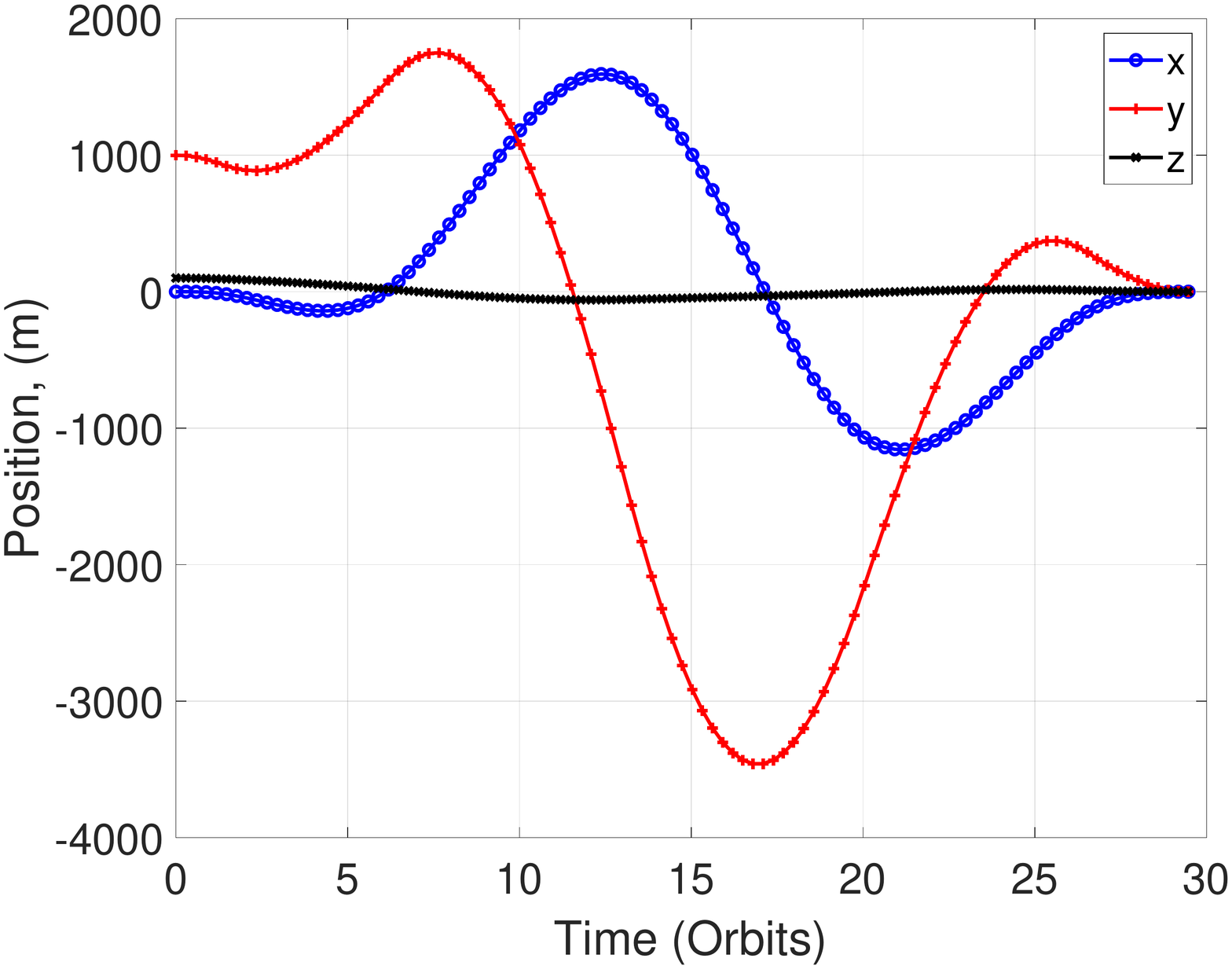}}
\subfigure[Velocity]
{\label{fig:33b}
\includegraphics[clip, trim=0cm 0cm 1cm 1cm, width=.48\textwidth]{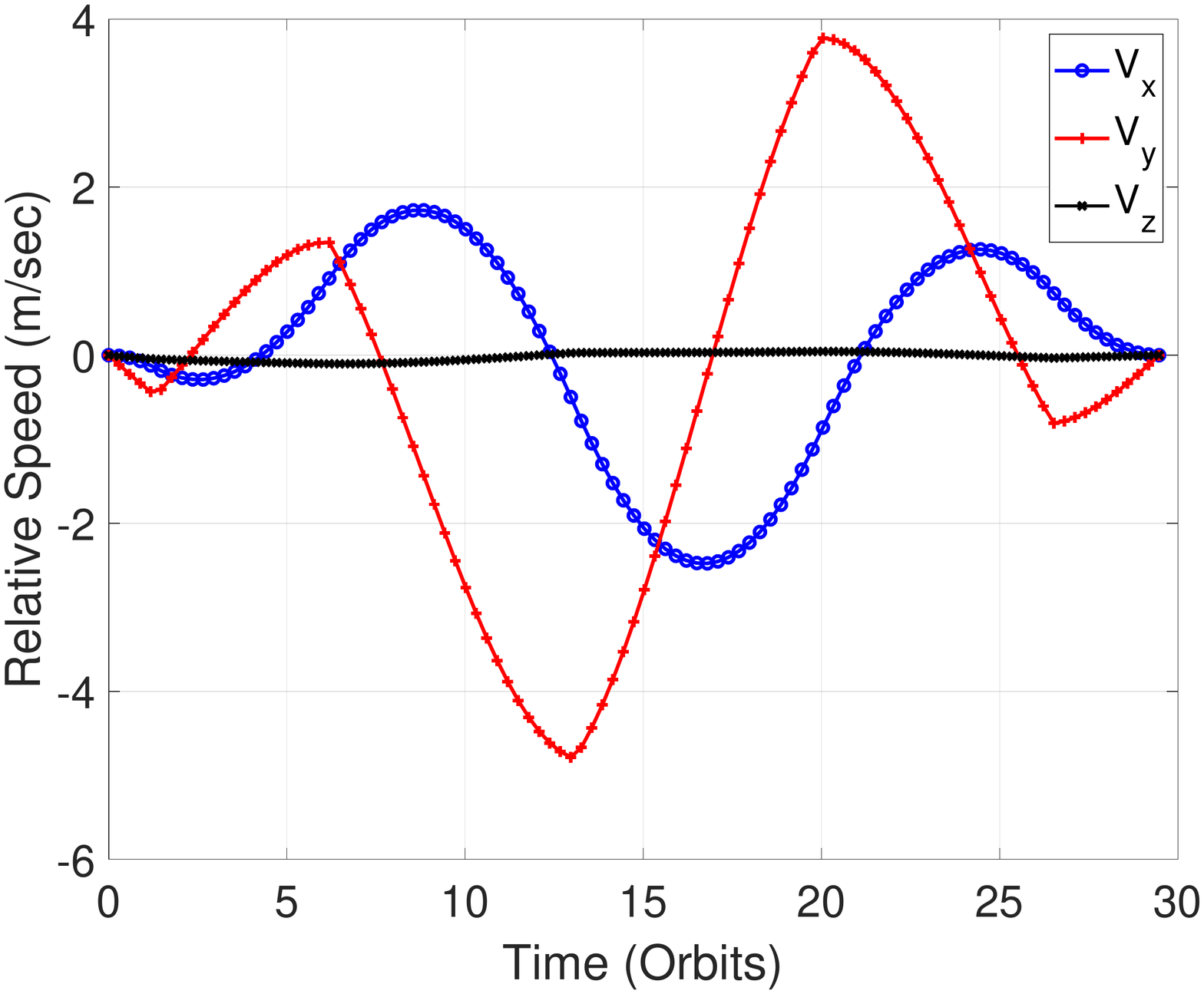}}
\caption{Error in Position and Velocity for Minimum Time Approach (Drag).}
\label{fig:33}
\end{figure}

\begin{figure}[tb]
\centering
{\includegraphics[clip, trim=0cm 0cm 0cm 0cm, width=.50 \textwidth]{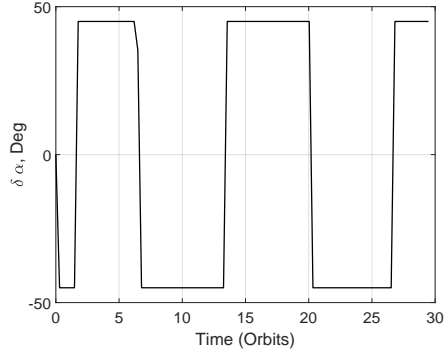}}
\caption{Control action for the Minimum-Time Problem (Drag).}
\label{fig:44}
\end{figure}
\noindent

\end{enumerate}

\subsection{Control with Atmospheric Drag and Lorentz Forces} \label{sec:5.2}
In this subsection, simulations are held between the optimal solutions with the combined control action of differential atmospheric drag and Lorentz forces for the whole trajectory of the radial-tangential-Normal rendezvous mission using two different approaches of minimum energy and minimum time. The weighting matrix ${R}$ is chosen as follows:
\begin{equation}\label{eq:costhybrid}
{R} =  \left[ \begin{array}{cc} 1&0\\0& 1.6 \times 10^7 \end{array} \right]
\end{equation}
\Cref{eq:costhybrid} renders the differential drag to be the dominant control action and the Lorentz as the minor auxiliary action to support the differential drag.
\begin{figure}[tb]
\centering
\subfigure[Position]
{\label{fig:6a}
\includegraphics[clip, trim=0cm 0cm 1cm 1cm, width=.48\textwidth]{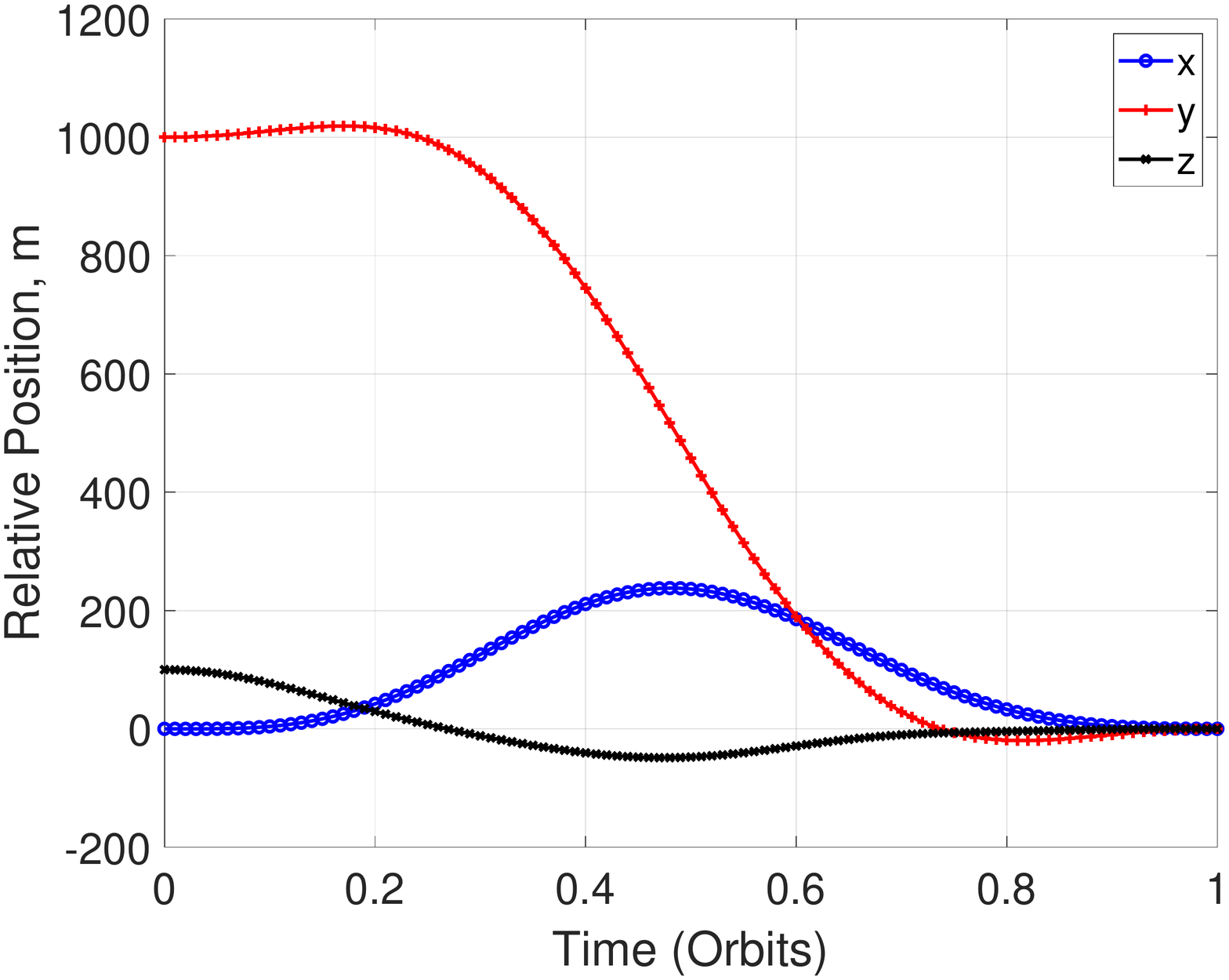}}
\subfigure[Velocity]
{\label{fig:6b}
\includegraphics[clip, trim=0cm 0cm 1cm 1cm, width=.48\textwidth]{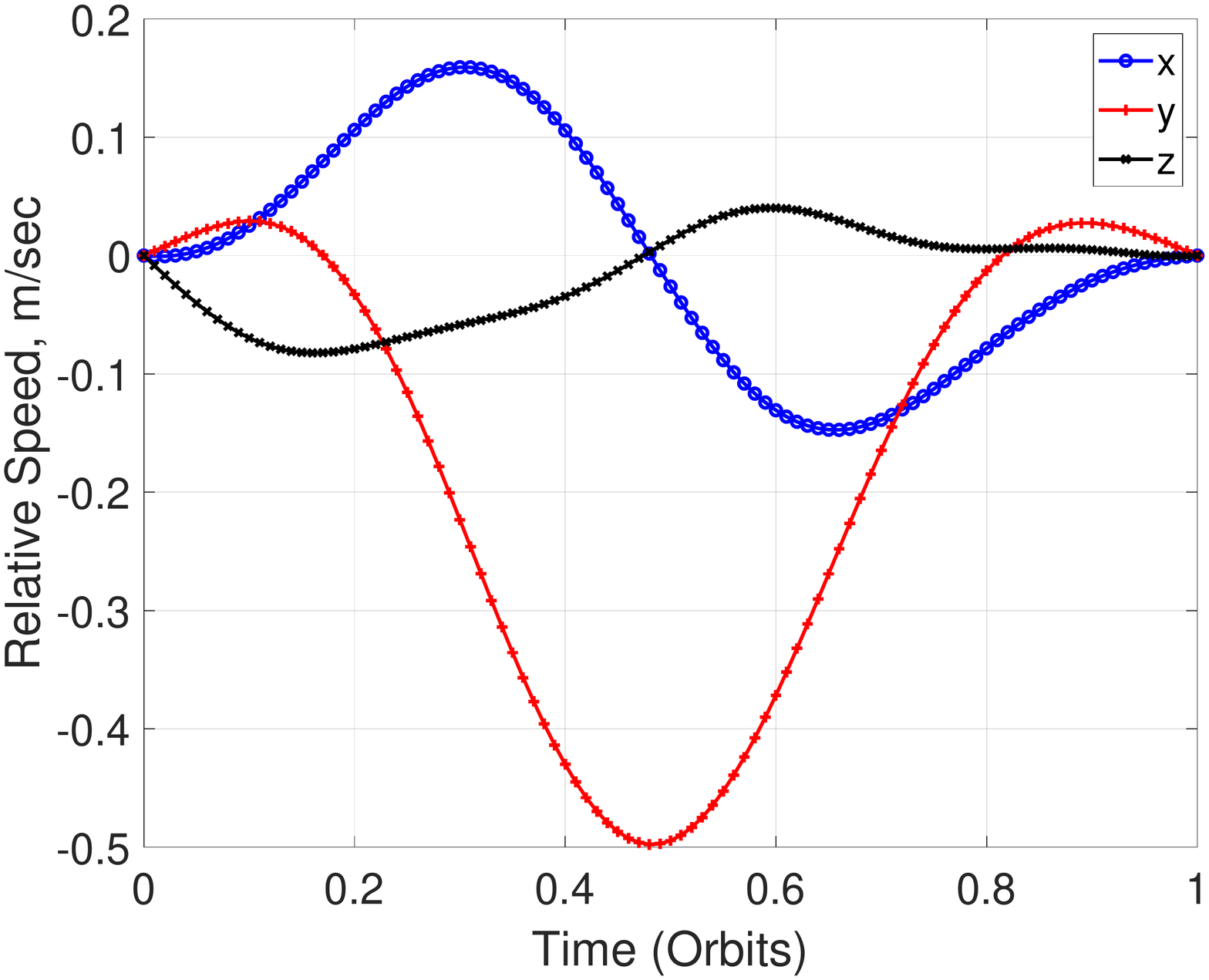}}
\caption{Error in Position and Velocity for Minimum Energy Approach (Drag + Lorentz).}
\label{fig:6}
\end{figure}

\begin{figure}[tb]
\centering
{\includegraphics[clip, trim=0cm 0cm 0cm 0cm, width=.50 \textwidth]{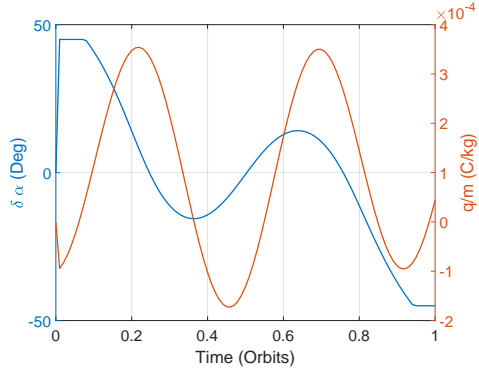}}
\caption{Control action For Minimum Energy Approach (Drag + Lorentz).}
\label{fig:7}
\end{figure}
\noindent

\begin{enumerate} 

\item Minimum Energy.

This part presents the results of the minimum-energy optimal control problem with combined differential atmospheric drag and Lorentz forces for the whole dynamics motion. \Cref{fig:6a} shows the time history of the relative position components, while \cref{fig:6b} presents the time history for the velocity components. It is shown that the Lorentz forces can handle the drawback of the required long duration to achieve this mission with differential drag alone.

In \cref{fig:7}, the trajectory of the control inputs $(\delta \alpha)$ and $q/m$ is presented with time $t_f= 6000$ sec, approximately equal to one orbital period. It is illustrated that the Lorentz charge over mass $q/m$ is extremely smaller than its saturation value while the drag plate angle is approaching its maximum value. This result indicates that the weighting matrix is properly chosen and that the mission can be achieved using atmospheric drag in less than one orbital period with the inconsiderable assistance of Lorentz forces with the total cost function for this mission $J = 1.11 \times 10^3$. 

\begin{figure}[tb]
\centering
\subfigure[Mean Control Actions]
{\label{fig:8a}
\includegraphics[clip, trim=0cm 0cm 1cm 1cm, width=.48\textwidth]{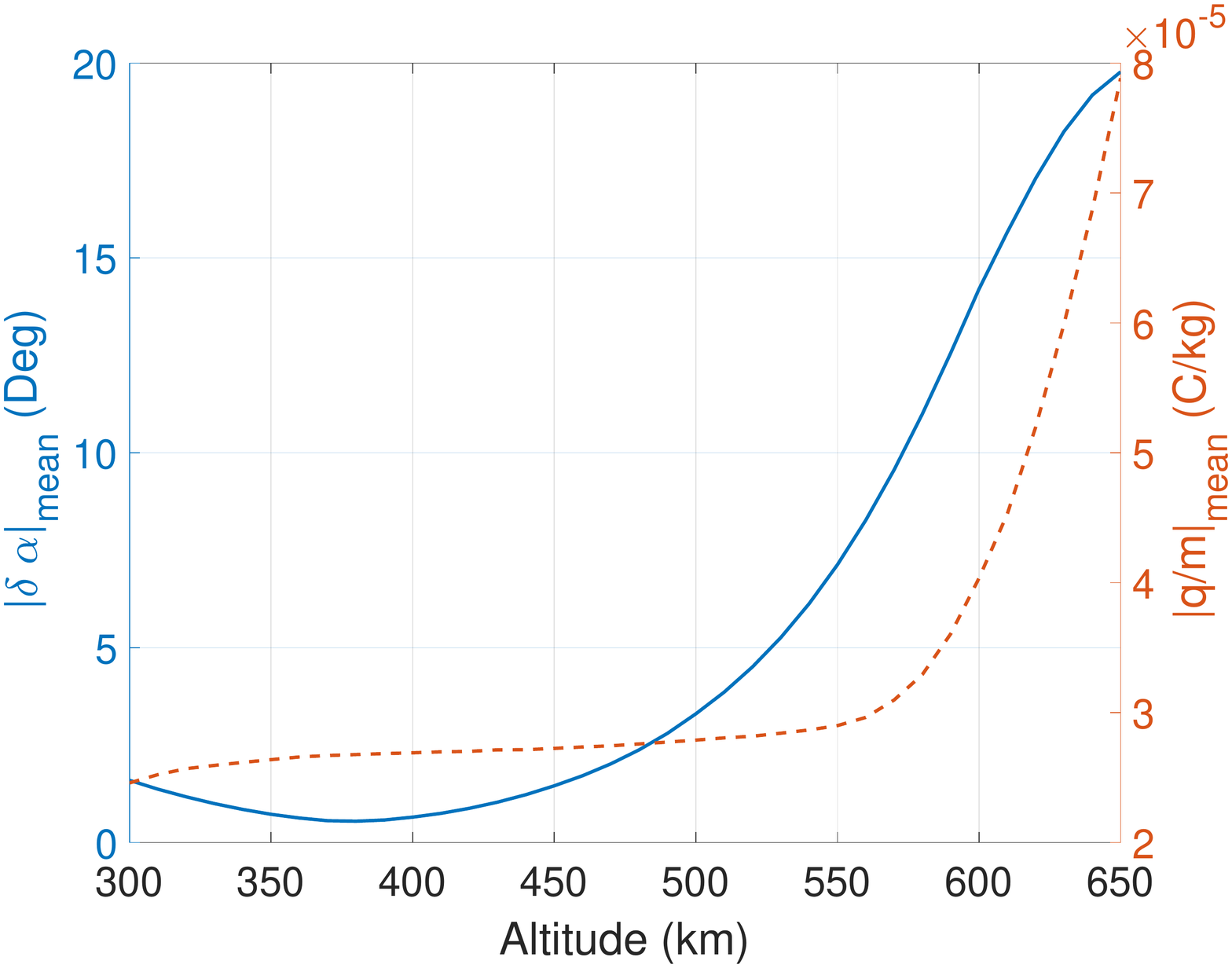}}
\subfigure[Total Cost Function]
{\label{fig:8b}
\includegraphics[clip, trim=0cm 0cm 1cm 1cm, width=.48\textwidth]{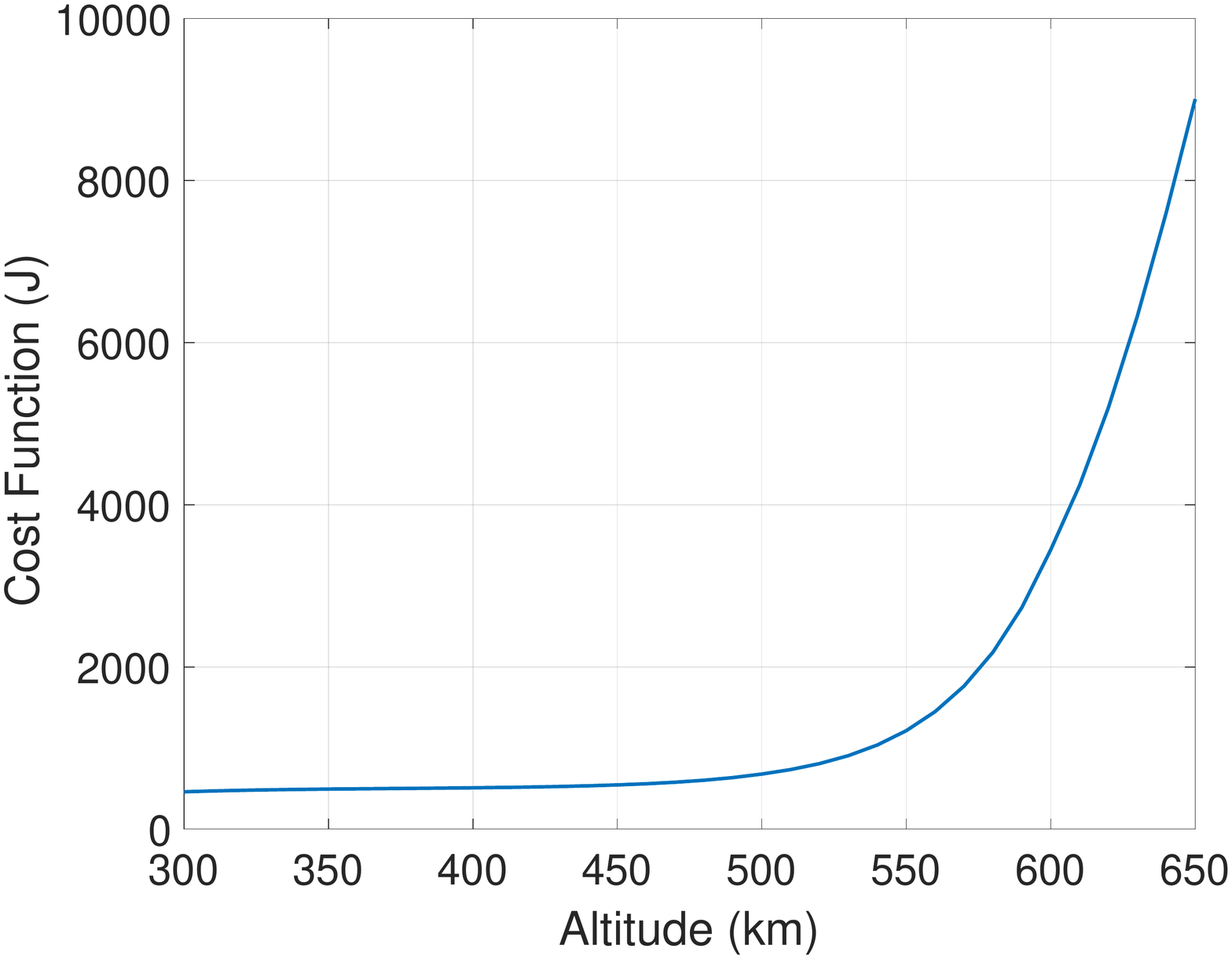}}
\caption{Mean Control Inputs and Total Cost Function with Various Altitudes $h_{ellp}$ (Drag + Lorentz).}
\label{fig:8}
\end{figure}

The results presented in \cref{fig:8} show the change in the mean control inputs and the cost function of the combined differential atmospheric drag and Lorentz forces for various altitudes from $h_{ellp}=300$ to $h_{ellp}= 650$ for the minimum-energy problem in five orbital periods. \Cref{fig:8a} illustrates that the mean control action of atmospheric drag and Lorentz forces changes smoothly with altitude. It is shown in \cref{fig:8b} that the cost function increases with a lower percentage than drag alone from  $J \approx 450$ for $h_{ellp} =300$ km to $J \approx 9000 $ for $h_{ellp} =650$ km.  

\item Minimum Time

\Cref{fig:9a} illustrates the time history for the position trajectory with the minimum-time approach, whereas \Cref{fig:9b} shows the values for the velocity components. We conclude from these figures that the minimum time required to satisfy the boundary conditions is nearly $0.52 $ orbital period, approximately equal to $3100$ sec.
The control inputs trajectories ($\delta \alpha(t)$ and $q(t)/m$) exhibit the bang-bang property, as shown in \cref{fig:10}.

\begin{figure}[tb]
\centering
\subfigure[Position]
{\label{fig:9a}
\includegraphics[clip, trim=0cm 0cm 1cm 1cm, width=.48\textwidth]{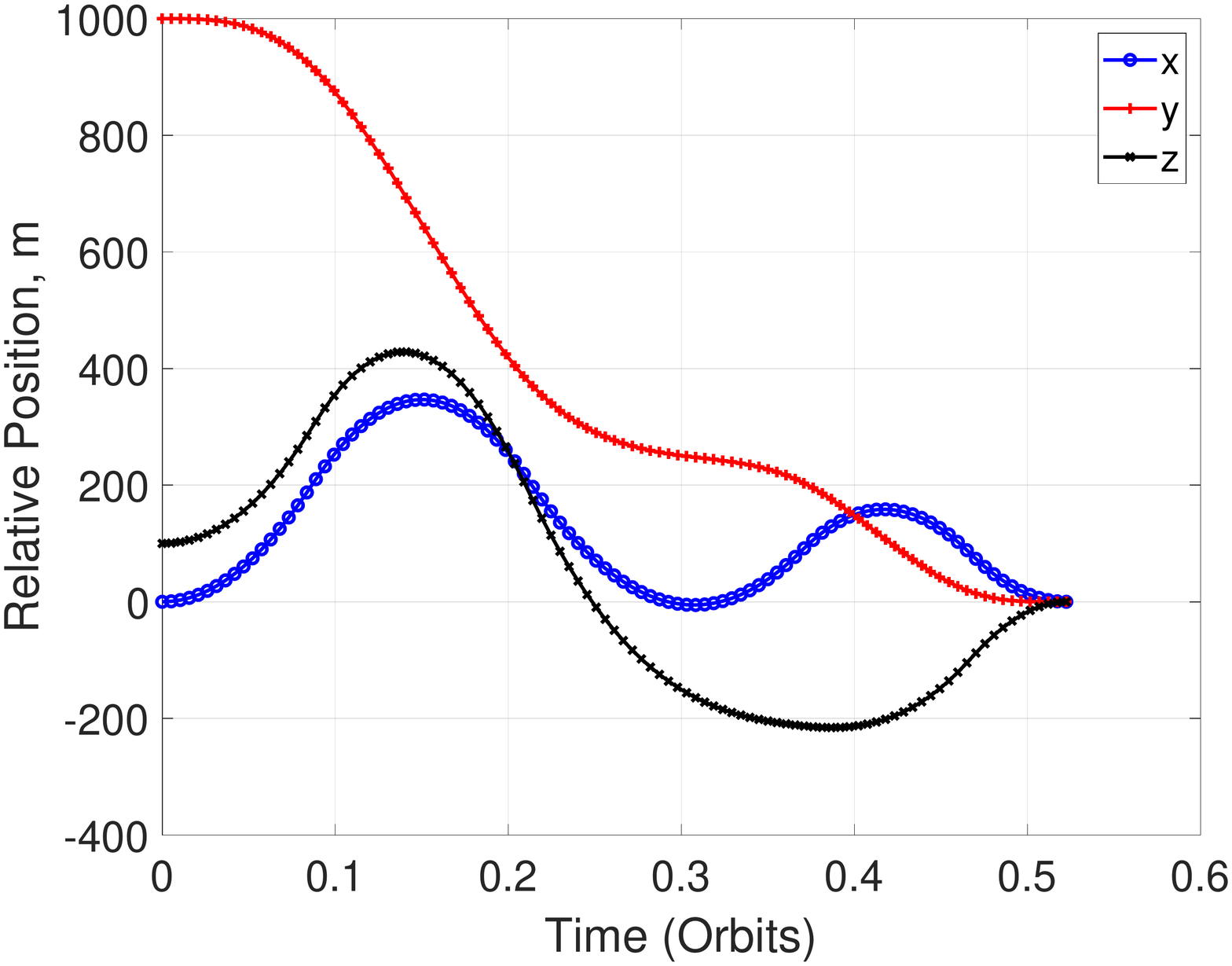}}
\subfigure[Velocity]
{\label{fig:9b}
\includegraphics[clip, trim=0cm 0cm 1cm 1cm, width=.48\textwidth]{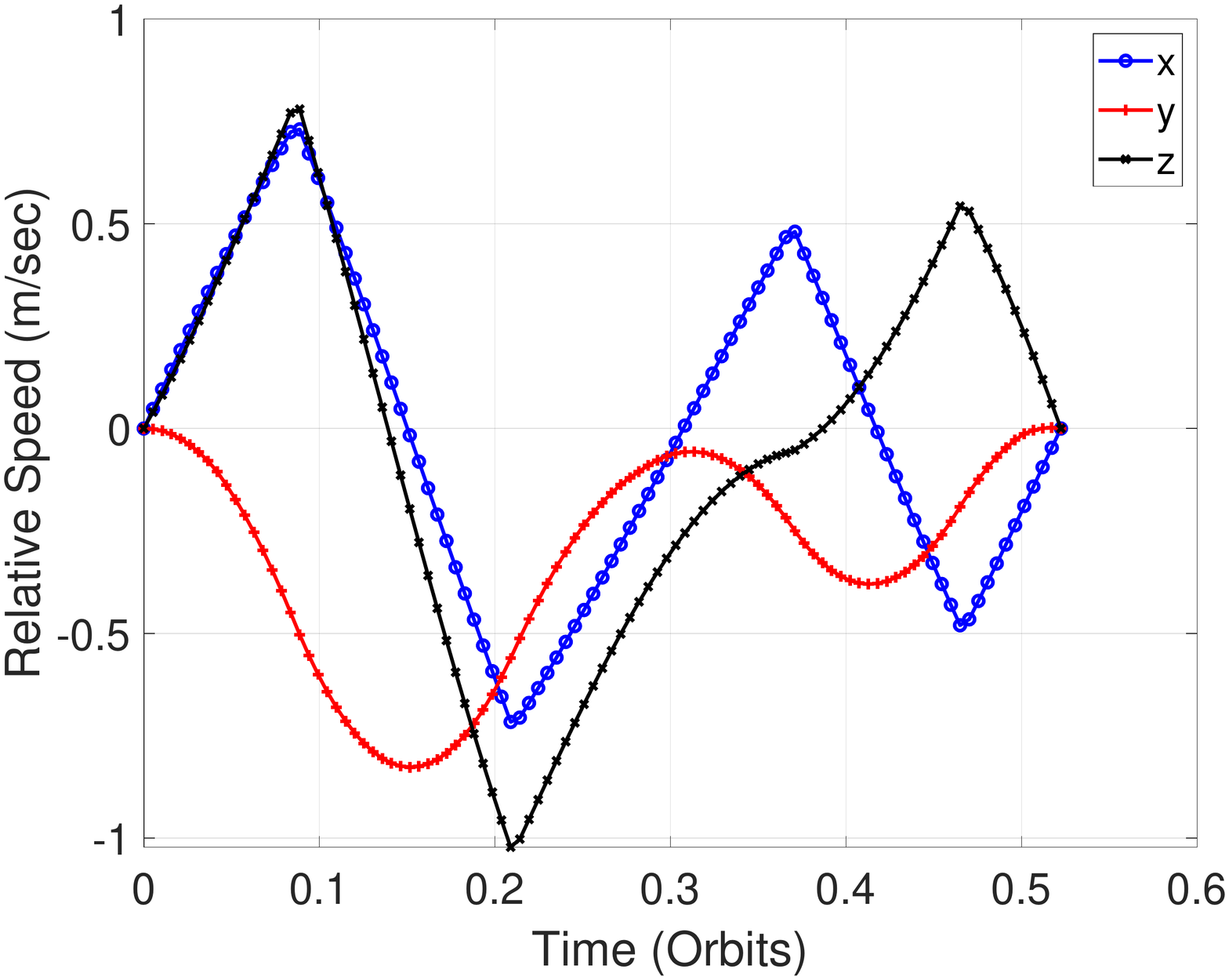}}
\caption{Error in Position and Velocity for Minimum Time Approach (Drag + Lorentz).}
\label{fig:9}
\end{figure}

\begin{figure}[tb]
\centering
{\includegraphics[clip, trim=0cm 0cm 0cm 0cm, width=.50 \textwidth]{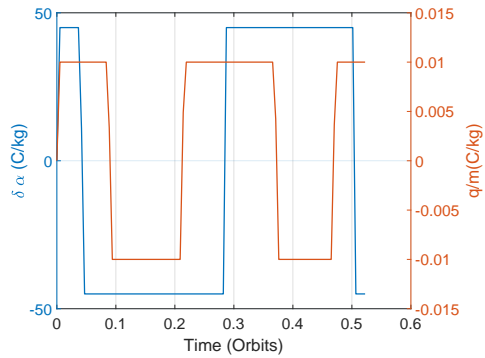}}
\caption{Control action for Minimum Time Approach (Drag +Lorentz).}
\label{fig:10}
\end{figure}
\noindent

It is illustrated that the integrated control action atmospheric drag and Lorentz forces can achieve the rendezvous mission in a short period (less than $0.52$ orbital period) with realistic parameters. The required time decreases to less than $2\%$ of its required time for the same mission with only the differential atmospheric drag. Therefore, it is concluded that using the hybrid control action of differential drag and Lorentz forces has high reliability in rendezvous missions in LEO.
\end{enumerate}

\section{Conclusions} \label{sec:6} 
This paper confirms space environmental forces' ability to control the nonlinear relative dynamics of Propellant-free rendezvous missions.
It illustrates that energy- and time-optimal trajectories can be obtained for the whole relative motion using the differential atmospheric drag alone. 
A detailed analysis of the impact of altitude above the Earth's surface indicated that the control actions and cost function increase significantly with altitude.
The atmospheric drag may only be practical for some rendezvous maneuvers as relatively long intervals are required, especially at high altitudes.
The Lorentz forces can support the differential drag even with inconsiderable charge amounts, increasing the possibility of obtaining feasible solutions for minimum-energy problems and shortening the required time excessively for minimum-time problems from several days to a few minutes. 
Future work will improve the control performance with the unmodeled dynamics by adding feedback control actions to cope with the generated trajectories.

\bibliographystyle{elsarticle-num}
\bibliography{Reference3}   % Use references. bib to resolve the labels.

\end{document}